\documentclass[mlabstract]{jmlr}

\usepackage{longtable}

\usepackage{booktabs}
\usepackage[load-configurations=version-1]{siunitx} 

\theorembodyfont{\upshape}
\theoremheaderfont{\scshape}
\theorempostheader{:}
\theoremsep{\newline}

\jmlrvolume{}
\firstpageno{1}

\jmlryear{2023}
\jmlrworkshop{Symmetry and Geometry in Neural Representations}

\title[Full-dimensional Characterisation of Single Stimulus-Response Distribution Geometries]{Full-dimensional Characterisation of Time-Warped\titlebreak Spike-Time Stimulus-Response Distribution Geometries}

 \author{\Name{James B. Isbister} \Email{james.isbister@epfl.ch}\\ 
 \addr The Blue Brain Project\\École Polytechnique Fédérale de Lausanne, Switzerland
 }

\begin{document}

\maketitle

\begin{abstract}
Characterising the representation of sensory stimuli in the brain is a fundamental scientific endeavor, which can illuminate principles of information coding. Most characterizations reduce the dimensionality of neural data by converting spike trains to firing rates or binned spike counts, applying explicitly named methods of ``dimensionality reduction'', or collapsing trial-to-trial variability. Characterisation of the full-dimensional geometry of timing-based representations may provide unexpected insights into how complex high-dimensional information is encoded. Recent research shows that the distribution of representations elicited over trials of a single stimulus can be geometrically characterized without the application of dimensionality reduction, maintaining the temporal spiking information of individual neurons in a cell assembly and illuminating rich geometric structure. We extend these results, showing that precise spike time patterns for larger cell assemblies are time-warped (i.e. stretched or compressed) on each trial. 
Moreover, by geometrically characterizing distributions of large spike time patterns, our analysis supports the hypothesis that the degree to which a spike time pattern is time-warped depends on the cortical area's background activity level on a single trial. Finally, we suggest that the proliferation of large electrophysiology datasets and the increasing concentration of ``neural geometrists'', creates ideal conditions for characterization of full-dimensional spike time representations, in complement to dimensionality reduction approaches.


\end{abstract}


\newpage
\section{Introduction}
\label{sec:intro}

Trial-to-trial variability implies that characterizing the neural representation of a single stimulus, requires characterization of the underlying distribution from which single trial representations are drawn. Whilst dimensionality reduction techniques have helped demystify aspects of neural processing \citep{gallego2017neural, humphries2020strong} we have become accustomed to low-dimensional perspectives. Characterisation of the full-dimensional geometry of timing-based representations may provide unexpected insights into how high-dimensional sensory and motor information is encoded. This could facilitate improved decoding and would constrain how cortical architecture parses information on a millisecond timescale. 




Recent research shows that the distribution of representations elicited over trials of a single stimulus can be geometrically characterized without the application of dimensionality reduction \citep{isbister2021clustering}, maintaining the temporal spiking information of individual neurons in a cell assembly and illuminating rich geometric structure. By studying responses to single whisker deflections in the barrel cortex, this work investigated the simplest correspondence between a piece of information and its cortical representation. Interestingly, single whisker deflections elicit only a single spike in the majority of responding neurons \citep{reyes2015laminar}, supporting the view that temporally \textit{atomic} stimuli and units of information are represented by short multi-neuron patterns of single spikes (Fig. \ref{fig:intro}A). In the visual cortex, such patterns even support visual discrimination \citep{resulaj2018first}. 

\begin{figure}[!b]
\floatconts{fig:intro}
{\caption{\textbf{Previous work.}
Reused with permission from \citet{isbister2021clustering}. \textbf{A:} Multi-neuron single spike pattern. \textbf{B:} First spike pair stimulus-response distribution with isolated correlated cluster. \textbf{C:} Cluster correlation angle different from $45^{\circ}$ implies modulation of spike times (heterogeneous) and spike time differences. \textbf{D:} State predictions correlated between clusters.}}
  {\includegraphics[width=1.0\linewidth]{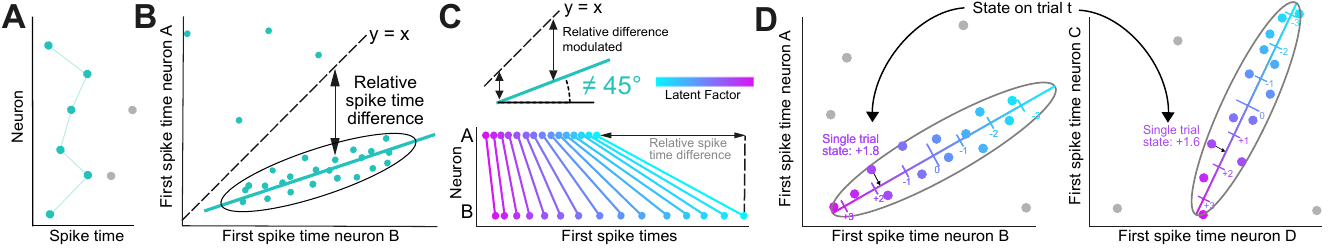}}
\end{figure}

Specifically, \citet{isbister2021clustering} studied how single spike patterns of two (and later four) cortical neurons co-represent a stimulus by plotting the first spike times of two neurons over repeated trials, and characterising the observed geometry (Fig. \ref{fig:intro}B). The study observed isolated regions (``clusters'') of high response probability, in which the spike times of the two neurons were positively correlated (i.e. the two neurons spiked earlier or later together on each trial). After controlling for stimulus-response adaptation, which can increase spike time latencies over repeated trials, the clusters remained positively correlated. This showed that the spike times of the two neurons, depended on a shared hidden variable, which varied from trial-to-trial. Interestingly, the correlation angle was often different from $45^{\circ}$, showing that the spike times of the two neurons were modulated differently by the hidden variable, such that the the timing difference between the spikes of the two neurons was also modulated (Fig. \ref{fig:intro}C). Finally, the study found that the variable modulating spike times was correlated between pairs of clusters (i.e. 4 neurons) and hypothesised that the cortical area's background activity level might correspond with the variable modulating spike times.

To test this hypothesis,
we explore whether a single dimensional hidden variable modulates the spike times of larger cell assemblies. In doing so, we geometrically characterize distributions of large spike time patterns for single stimuli. This demonstrates that multi-neuron single spike patterns, which could be the building blocks of neural representation, provide a tractable opportunity to characterize full-dimensional neural representations. Moreover, trial-to-trial variability in spike time patterns and dimensionality-reduced representations might be governed by the same low-dimensional mechanisms, and could explain why millisecond-precise spike time patterns are challenging to detect \citep{russo2017cell, stella20193d}.
Whilst biases towards explaining task variables promote dimensionality reduction approaches (see Discussion), we suggest that the increasing concentration of ``neural geometrists'' and electrophysiology datasets, creates ideal conditions for complementary characterization of full-dimensional spike time representations. 


\section{Results}

Data analysed here and in \citet{isbister2021clustering} was reused from \citet{reyes2015laminar}. Stimuli were 2ms single whisker deflections with each stimulus condition defined by the identity of the stimulated whisker and the inter-trial-interval, with 80-200 trials made per condition. Spike sorting was applied to recordings from eight 16 electrode shanks spanning layers 2/3, 4 and 5, and separated across three to four cortical barrels in anaesthetized rats.

\citet{isbister2021clustering} developed a clustering pipeline that detected isolated regions of high-response probability (``clusters'') in first spike response distributions for neuron pairs. The algorithm was validated to prevent detection of spurious clusters. If the spike times of cluster samples were modulated by stimulus-response adaptation, the adaptative trends were removed. Fig. \ref{fig:results}A visualizes the clusters extracted for a single condition after control for adaptation. All 12 clusters are positively correlated (many with correlation angles different from $45^\circ$), with 8 neurons \textit{participating} in at least one cluster. Supplementary Video 1 shows figures for all experimental conditions.

Here we test whether a shared single dimensional hidden variable is responsible for observed spike time co-variability across larger cell assemblies, as this would support the dependence of trial-to-trial variability on the background activity level. Firstly in support of this, the single trial states (i.e. the global background activity level) predicted by each cluster are correlated between clusters (Fig. \ref{fig:results}B). Secondly, using the spike times from clusters we can calculate the correlation matrix between participating neurons, including for neuron pairs for which a cluster was not originally detected (Fig. \ref{fig:results}C, left). A large proportion of neuron pairs show significant correlations ($p < 0.05$; Fig. \ref{fig:results}C, right). Moreover, almost all of the neuron pairs are positively correlated; even neuron pairs which were not shown to be significantly correlated (Fig. \ref{fig:results}C), supporting low-dimensional modulation. These findings were general over conditions (Fig. \ref{fig:results}D). Importantly, for the majority of conditions, only the 1st and 2nd eigenvalues of factor analysis were significant based on the Kaiser criterion ($>1$), with the 1st eigenvalue strongly explaining variance (Fig. \ref{fig:results}E).

\begin{figure}[!t]
\floatconts{fig:results}
{\caption{\textbf{New preliminary results.} \textbf{A:} For an example condition, cluster spike times coloured by the estimated cortical state, calculated as the value of the first principle component of the neuron pair cluster spike times. \textbf{B:} For an example condition, coloured points show cortical states predicted from each detected cluster, by trial. Black line shows the mean on each trial. \textbf{C:} Left: correlation matrix between cluster spike times of different neuron pairs. Right: P-values of the same correlations. Black if $>0.05$. \textbf{D:} R and P-value histograms of data as described in C but over all conditions. \textbf{E:} Eigenvalue by factor analysis eigenvector. \textbf{F:} R-values of correlation between singe trial cortical state and first spike times. \textbf{G:}  Spike time of each neuron coloured by the global estimate of cortical state on each trial. Lines show means of 10 highest state trials and 10 lowest state trials.}}
  {\includegraphics[width=1.0\linewidth]{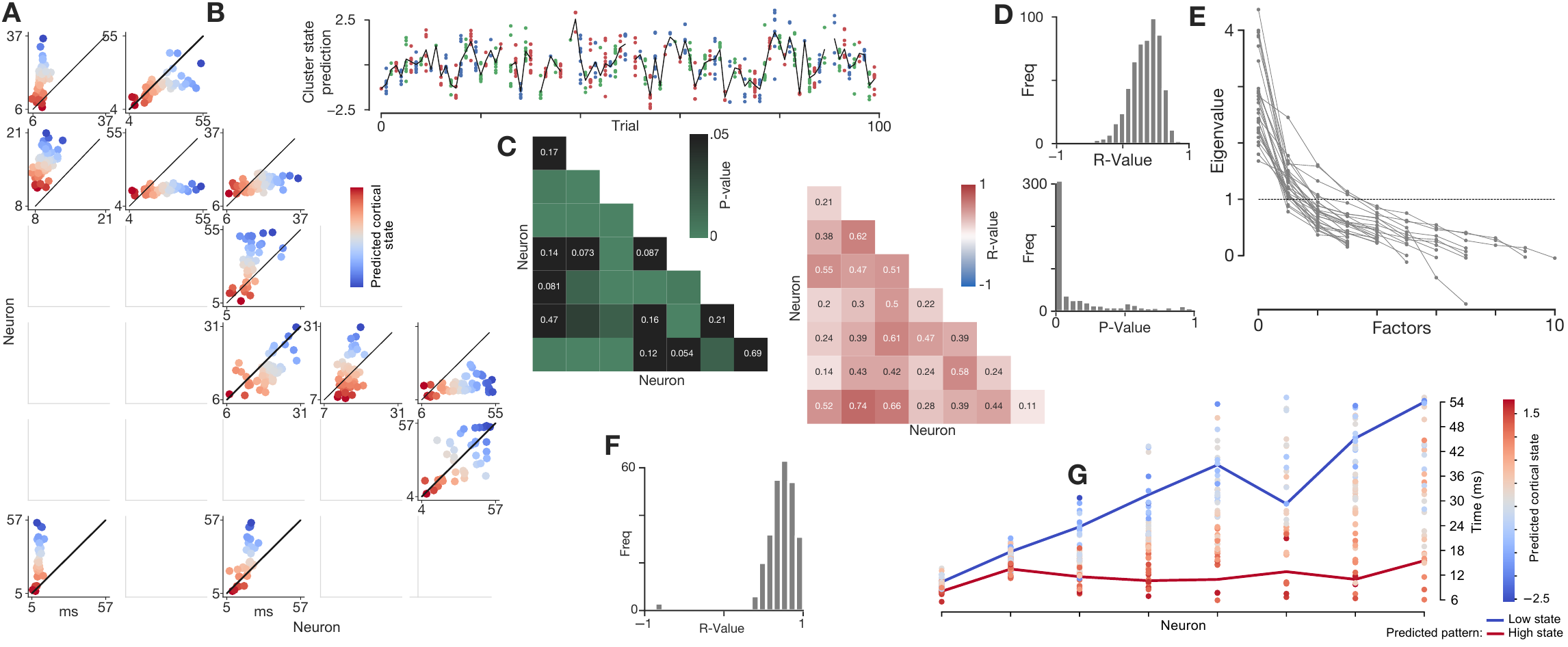}}
\end{figure}




These results motivate a single dimensional ``global cortical state'' estimate: the mean of single cluster cortical state estimates (Fig. \ref{fig:results}B). 
This estimate explained a large amount of variance in cluster first spike times (Fig. \ref{fig:results}F); future work will employ a leave one out analysis. We next plot the spike time of each neuron coloured by the single trial global cortical state estimate, illustrating heterogeneous modulation of single neuron spike times by the global state (Fig. \ref{fig:results}G). The visualisation shows the average response for the 10 highest state trials vs the 10 lowest state trials, highlighting how the large multi-neuron single spike patterns is dramatically modulated by the shared latent factor. 

These results suggest that the response distribution is highly structured and can be characterized as a high-dimensional high-response probability cluster surrounded by currently unexplained noise (or poorly sampled regions). The cluster is positively correlated due to a single dimensional latent factor oriented away from the unitary line. The cluster can be approximated by a Gaussian. For each single trial, the response distribution has missing values for different neurons due to low response reliability. As a first approximation, the response distribution was characterized by finding clusters in two-dimensional slices of the distribution, and analyzing the hidden variables modulating clusters in different slices. 


\section{Discussion}
\label{sec:discussion}

To our knowledge, such a precise geometrical characterisation has never previously been made. Low-dimensional modulation of larger multi-neuron single spike patterns strengthens the hypothesis of \citet{isbister2021clustering} that modulation reflects low-dimensional latent factors observed in dimensionality-reduced representations \citep{kao2015single, pandarinath2018inferring}. The single trial global cortical state estimate can thus be compared with trial-to-trial variability in reduced-dimensions. The metric may also correlate with the spiking activity of other neurons (not just cluster neurons), as well as secondary and tertiary spikes. As spike times are modulated positively, the biophysical nature of such low-dimensional modulation is likely to correspond with fluctuations in ``shared excitability'', which also modulate spike counts \citep{lin2015nature}. Spatially, this may correspond with travelling waves, indirectly measurable through the LFP \citep{davis2020spontaneous}.

The characterized representations may aid in understanding neural coding, such as theoretical encoding capacity or how neurons integrate temporal patterns. Relative timing differences between spikes must be considered, which are interpretable downstream. \citet{isbister2021clustering} previously suggested that low-dimensional modulation over multiple cortical layers and columns makes it possible that downstream neurons are also modulated by the same fluctuations introducing the possibility that downstream neurons can decode patterns conditioned on cortical state. Estimates of state-dependent coding capacity were made by \citet{isbister2021clustering}, and will likely be improved using the estimated global cortical state.

The precise characterisation of multi-neuron first spike response distributions comes nearly 100 years after the discovery of the action potential \citep{adrian1926impulses}, which were described as ``scarcely more complex than a succession of dots in Morse Code'' \citep{adrian1932mechanism}. In addition to the challenges of multi-electrode recording, a number of factors perhaps explain why such a characterization was long overlooked. Rather than studying the fundamental correspondence between a simple stimulus and its stochastic neural representation, researchers and funders are often biased towards demonstrating encoding of features or task relevant information. As such, duration limited experimental sessions often favor larger numbers of conditions over high trial counts per condition. Analyses are then applied which work robustly with low trial counts, in exchange for temporal granularity. In a self-proliferating cycle, statistical techniques and experiments are then developed in the same vein. Task variables are often low-dimensional, due to the complexity of training animals, and the need to create controlled and interpretable paradigms. This leads to low-dimensional modulation of population activity by task variables, which again proliferates dimensionality-reduction. In sensory modalities such as vision, the brain must rapidly create high-dimensional representations, however.

Openly available, large biophysically-detailed models are becoming gradually more abundant \citep{billeh2020systematic, isbister2023modeling}. A recent cortical model precisely captures layer-wise population response to whisker deflections, and selective propagation of activity to downstream areas \citep{isbister2023modeling}. Such a model provides the opportunity to generate unlimited samples from stimulus-response distributions, whilst controlling nonstationarities. As the model is refined, it is likely that the response distributions will better reflect those characterized \textit{in vivo}, and the model can be used to study the propagation of activity through cortical architecture at full spatiotemporal scale.

\subsection*{Funding}
This study was supported by funding to the Blue Brain Project, a research center of the École polytechnique fédérale de Lausanne (EPFL), from the Swiss government’s ETH Board of the Swiss Federal Institutes of Technology.

\section*{Supplementary Video 1}
Supplementary Video 1 is available at: \href{https://t.ly/iOhhX}{t.ly/iOhhX}

\bibliography{pmlr-sample}
\end{document}